\documentclass[journal,onecolumn]{IEEEtran}
\IEEEoverridecommandlockouts
\usepackage{algorithm}
\usepackage{algorithmicx}
\usepackage{algpseudocode}
\usepackage[latin1]{inputenc}
\usepackage[cmex10]{amsmath}
\usepackage{amsfonts}
\usepackage{amssymb}
\usepackage{graphicx}
\usepackage{verbatim}
\usepackage{array}
\usepackage{multirow}
\usepackage{dcolumn}
\usepackage{xcolor}
\usepackage{url}
\usepackage{graphicx}
\DeclareGraphicsExtensions{.eps}
\usepackage{balance}

\usepackage[noadjust]{cite}
\usepackage{bbm}
\usepackage{float} 

\usepackage{mathrsfs}
\usepackage{color}
\usepackage{booktabs}
\usepackage{changes}

\newlength\myindent
\renewcommand{\algorithmicrequire}{\textbf{Input:}}
\renewcommand{\algorithmicensure}{\textbf{Output:}}

\newcommand{\ceiling}[1]{\lceil#1\rceil}
\newcommand{\ceilingFrac}[1]{\bigg\lceil#1\bigg\rceil}
\newcommand{\ceilingtFrac}[1]{\big\lceil#1\big\rceil}

\begin{document}
%

\title{FPGA-Embedded Linearized Bregman Iterations Algorithm for Trend Break Detection}

%
\author{Felipe~Calliari,~Gustavo~C.~Amaral,~Michael~Lunglmayr~\IEEEmembership{Member,~IEEE}
\thanks{}
\thanks{F.~Calliari is with the Center for Telecommunications Studies, Pontifical Catholic University of Rio de Janeiro, RJ, Brazil (e-mail: felipe.calliari@opto.cetuc.puc-rio.br).}
\thanks{M.~Lunglmayr is with the Institute of Signal Processing, Johannes Kepler University, Linz, Austria (e-mail: michael.lunglmayr@jku.at).}
\thanks{G.~C.~Amaral is with the Center for Telecommunications Studies, Pontifical Catholic University of Rio de Janeiro, RJ, Brazil and with the QC2DLab, Kavli Foundation, Technical University of Delft, The Netherlands (e-mail: gustavo@opto.cetuc.puc-rio.br).}
\thanks{Copyright (c) 2017 IEEE. Personal use of this material is permitted.  However, permission to use this material for any other purposes must be obtained from the IEEE by sending a request to pubs-permissions@ieee.org.}
}


\maketitle
\begin{abstract}
Detection of level shifts in a noisy signal, or trend break detection, is a problem that appears in several research fields, from biophysics to optics and economics. Although many algorithms have been developed to deal with such problem, accurate and low-complexity trend break detection is still an active topic of research. The linearized Bregman Iterations have been recently presented as a low-complexity and computationally-efficient algorithm to tackle this problem, with a formidable structure that could benefit immensely from hardware implementation. In this work, a hardware architecture of the Linearized Bregman Iterations algorithm is presented and tested on a Field Programmable Gate Array (FPGA). The hardware is synthesized in different sized FPGAs and the percentage of used hardware as well as the maximum frequency enabled by the design indicate that an approximately 100 gain factor in processing time, with respect to the software implementation, can be achieved. This represents a tremendous advantage in using a dedicated unit for trend break detection applications.
\end{abstract}

%

\IEEEpeerreviewmaketitle

\section{Introduction}
\label{sect:1}
\IEEEPARstart{T}{rend break} detection in the presence of noise is a broad problem that can be found across different research fields \cite{LunglmayrTIM2018, basseville1983desgin, rabiner1989tutorial, soutoSAM2016, lorenz2014sparse}. For that reason, several different methodologies have been proposed in the literature \cite{rea2010identification, storath2014jump, WeidJLT2016}, with the ones that make use of $\ell_1$ regularization to counter the problem's inherent high-dimensionality arguably figuring as the most successful ones \cite{kim2009ell_1, lunglmayr2016efficient}. Such an approach is required for highly reliable estimation results \cite{WeidJLT2016}. Even though such regularization allows the problem to be solved in a computationally efficient manner (usually associated to a complexity which is proportional to a polynomial function of the number of inputs), the fact that a computer can solve the problem does not necessarily mean that the result is achieved quickly, practically speaking. In certain contexts, achieving elapsed algorithm times in the order of seconds as opposed to minutes may yield a substantial impact on the application \cite{calliariPhotoptics2018, herrera2016high}.

It is a widespread notion that certain problems, despite their complexity, may be accelerated depending on the implementation; parallel programming, in which several parts of the same procedure are processed independently and simultaneously, is one of the most celebrated examples \cite{rumelhart1986parallel}. Field Programmable Gate Arrays (FPGAs) are extremely versatile hardware structures that offer \cite{xie2017efficient, greisen2013evaluation, cong2001design}: great flexibility to design high speed high-density digital hardware; easiness of programability and reconfiguration; energy efficiency; high resource utilization; low cost; and the possibility to combine parallel processing structures with serial control units. FPGAs haven been used as a versatile computing platform accelerating algorithms through dedicated and carefully designed architectures in a wide range of fields such as cryptography \cite{CryptoFPGA}, image processing\cite{ImageFPGA}, and machine learning \cite{MachineLearnFPGA}. Recently, linearized Bregman Iterations (LBI), a class of implementation-efficient and low-complexity algorithms, has been presented as an extremely attractive solution for trend break detection \cite{LunglmayrTIM2018}. In particular, both the structure of the trend break detection problem and of the LBI algorithm's allow for simple hardware units, relying mainly on adders and efficient memory management, to conduct the core procedure, thereby avoiding hardware-complex multiplication and division operations \cite{meyer2004digital}.

In this work, the hardware implementation of the LBI algorithm is studied in depth and is simulated and synthesized for different FPGAs. A novel hardware architecture is presented and its main processing units are discussed. VHDL simulation environments enables a step-by-step comparison and validation of the processing stages referenced by the computer algorithm implementation \cite{LunglmayrTIM2018}. Hardware synthesis results allow determination of both device usage with different FPGA sizes and maximum clock frequency; the former, combined with the average number of clock cycles per iteration loop, make total processing time calculation possible for different problem instance sizes. A reduction factor on the elapsed algorithm time of approximately 100 is achieved, which represents a substantial upgrade and warrants usage of dedicated hardware for trend break detection.

Due to the limited memory resources available in an embedded processing unit as opposed to a standard personal computer, different data formatting was employed in the hardware implementation. This not only allows for increased data length manipulation inside the embedded unit (with estimated $\geq60000$ data points for a mid-sized FPGA), but also avoids the usage of complex and often slower arithmetic units to handle floating point representation \cite{hormigo2017hub}. The comparison between results using the 20-bit fixed-point (\textit{SFIXED}) \cite{sfixedPackage} format and standard 64-bit double representation have been conducted in this work and negligible discrepancies were observed.

The paper is divided as follows. In Section \ref{sect:2}, a brief review of the LBI algorithm for trend break detection is performed, including the structure of the candidate matrix and the pseudocode based on which the hardware architecture is developed. Section \ref{sect:3} presents the digital hardware architectural concept as well as focused descriptions of its main units; the estimated number of clock cycles until the algorithm elapses is derived based on this architecture. In Section \ref{sect:4}, comparative results between the simulated hardware implementation and the Julia code of the LBI are discussed. Synthesis parameters for two target  FPGAs (\textsc{Altera Cyclone V} and \textsc{Altera Stratix V}) are also reported. Case studies and implementation results are discussed in Section \ref{sect:5}, and Section \ref{sect:6} concludes the paper.

\section{The Linearized Bregman Iterations Algorithm for Trend Break Detection}
\label{sect:2}
Under the assumption that the trend break detection problem is a sparse one, i.e., the number of candidate vectors that describe the signal of interest is much smaller than the number of observations, it can be cast into the combined $\ell_1/\ell_2$ problem of the form \cite{LunglmayrTIM2018}:
\begin{equation}
\min_{\boldsymbol{\beta}} \lambda ||\boldsymbol{\beta}||_1 + \frac{1}{2}||\boldsymbol{\beta}||_2^2 \hspace{0.2cm} s.t. \hspace{0.2cm} {\bf A} \boldsymbol{\beta}={\bf y},
\label{eq:linearizedBregman}
\end{equation}
where ${\bf A}$ is the dictionary, with each candidate vector stored in a column, $\boldsymbol{\beta}$ is the vector containing the coefficients of the weighted linear combination of dictionary vectors that will approximate the signal of interest represented by the data series ${\bf y}$, and $\lambda$ is a parameter that adjusts the weight of the $\ell_1$ versus the $\ell_2$ norm. Adaptation of the Linearized Bregman Iterations algorithm to trend break detection has been presented in \cite{LunglmayrTIM2018} in a context where a linear trend is also expected in the signal of interest. In order to simplify and generalize the implementation, this linear trend is not considered in the current implementation. Incorporating the linear trend in the proposed architecture can, however, be easily done.

Throughout the manuscript, the length, in data points, of the signal of interest ${\bf y}$ will be defined as $N$, i.e., ${\bf y}$ and $\boldsymbol{\beta}$ are N-dimensional vectors and ${\bf A}$ is an $N\times N$ matrix. The Linearized Bregman Iterations algorithm has a periodic structure, involving, in a single iteration, an approximate gradient descent (AGD) followed by a non-linear shrink function of the form: $\textrm{shrink}\left(v,\lambda\right)= \max\left(|v|-\lambda,0\right)\cdot\textrm{sign}\left(v\right)$ \cite{LunglmScaled}. Due to the special structure of the candidate dictionary matrix $\bf{A}$ for the trend break detection problem, namely:
\begin{equation}
{\bf A} = \begin{bmatrix}
1        & 0         & 0       & \cdots & 0         & 0      \\
1        & 1         & 0       & \cdots & 0         & 0      \\
1        & 1         & 1       & \cdots & 0         & 0      \\
\vdots   & \vdots    & \vdots  & \ddots & \vdots    & \vdots \\
1        & 1         & 1       & \cdots & 1         & 0      \\
1        & 1         & 1       & \cdots & 1         & 1
\end{bmatrix},
\label{eq:matrixX}
\end{equation} its storage is not necessary for the AGD calculation, as the latter can be rewritten as
\begin{align}
\begin{aligned}
\textbf{v}^{\left(i+1\right)} &= \textbf{v}^{\left(i\right)} + \frac{\textbf{a}_k}{||\textbf{a}_k||^2_2}\left(y_k-\textbf{a}_k^\textrm{T}\boldsymbol{\beta}^{\left(i\right)}\right) \\
&= \textbf{v}^{\left(i\right)} + \frac{\textbf{a}_k}{||\textbf{a}_k||^2_2}\left(y_k-\sum_{s=1}^{k+1}\beta_s\right)
\end{aligned}
\end{align}
where the $\textbf{a}_k$ represent columns of the candidate matrix, the superscripted $i$ represents the iteration index, and the index $k\in\left[1:N\right]$ controls the cyclic re-use of rows of ${\bf A}$ as the iteration index evolves, i.e., $k =  \textrm{mod} \left(\left(i-1\right), N\right) + 1$.

The $\textbf{a}_k$, in turn, have an interesting structure that allows the AGD to be further optimized and the calculation to be performed only for those indices where $a_{k,j}\neq0$. In other words (and also considering the fact that $||\textbf{a}_i||_2^2=k$),
\begin{align}
v_j^{\left(i+1\right)} = \left\{
\begin{array}{ll}
v_j^{\left(i\right)} + \frac{1}{k}\left(y_k-\sum_{s=1}^{k+1}\beta_s\right) ,& a_{k,j} = 1\\
v_j^{\left(i\right)} ,& a_{k,j} = 0
\end{array}
\right.,
\end{align}
which, considering computational implementation, translates into accessing and manipulating only those values of vector $\textbf{v}^{\left(k\right)}$ up to index $j$. A final observation of the structure of matrix ${\bf A}$ (namely, the fact that it is a square matrix) reveals that a single index $k$ is sufficient to control an iteration of the algorithm. The resulting procedure, presented as a pseudocode in Algorithm \ref{alg:linBreg}, efficiently solves the trend break detection problem with low memory usage. 

\begin{algorithm}[ht]
\caption{Linearized Bregman Iteration for Trend Break Detection}
\label{alg:linBreg} 
\algorithmicrequire{ Measurement vector $\boldsymbol{y}$, $\lambda$, $\boldsymbol{\beta}_\text{start}$, $\boldsymbol{v}_\text{start}$, $L$}\\
\algorithmicensure{ Estimated $\hat{\boldsymbol{\beta}}$}
\begin{algorithmic}[1]
\small
\State{$\boldsymbol{\beta}^{\left(0\right)}\leftarrow \boldsymbol{\beta}_\text{start}$}
\State{${\bf v}^{\left(0\right)}\leftarrow \boldsymbol{v}_\text{start}$}
\State{$i \gets 1$}
\While{$i<L$} 
\State $k \leftarrow \text{ mod }( (i-1), N ) + 1 $  \Comment {cyclic re-use of rows of ${\bf A}$}
\State $\mu_k \leftarrow  \frac{1}{k}$
\State $e \leftarrow \left(y_k - \sum_{s=1}^{k+1} \beta_s^{\left(i\right)}\right)$  
\Statex \Comment {instantaneous error with inner product}
\State $d \leftarrow \mu_k e$
\For{$j=1..k$}
\State{${v}_j^{\left(i+1\right)}\leftarrow v_j^{\left(i\right)}+ d$}
\State{$\bf{\beta}_j^{\left(i+1\right)}\leftarrow$ shrink $\left(v_j^{\left(i+1\right)},\lambda\right)$}
\EndFor
\State $i \leftarrow i +1$
\EndWhile
\end{algorithmic}
\end{algorithm} 

It is important to note that only the computation-heavy part of the algorithm is depicted in the pseudocode of Algorithm \ref{alg:linBreg}. Its purpose is to identify the relevant non-zero values of the $\hat{\boldsymbol{\beta}}$ vector that compose the output or, in other words, reduce the dimension of the detection space focusing on the subspace spanned by the relevant candidate vectors. After this procedure, it is usual to perform an Ordinary Least Square (OLS) in this reduced subspace in order to remove any biasing introduced by the algorithm; operating on the reduced subspace found by the LBI drastically reduces the complexity of the OLS. This step, which involves matrices transposing and inverting, can be efficiently conducted in a standard personal computer and, even though this could also be implemented in the same hardware structure that contains the core algorithm \cite{LunglmayrTIM2018}, the goal of this work is to present the latter and the OLS step is left as a post-processing to be performed in a different processing unit.

Also left as a pre-processing step is the scaling of the data vector ${\bf y}$, which is necessary to ensure the correct behaviour of step 7 in Algorithm \ref{alg:linBreg} when using the 20-bit fixed-point format; namely, that no overflow of the arithmetic dynamic range is observed when performing the summation of $\beta$ values. The scaling is intimately connected to the available arithmetical dynamic range, which, in turn, is connected to the memory resources of the FPGA board, thereby constituting a design-related compromise relationship: in case of over-scaling, the arithmetic dynamic range will be hindered; to overcome this, a higher number of bits can be assigned to the data points, which then increases the resources necessary in the FPGA board; on the other hand, in case of under-scaling, the results may overflow, creating errors that can jeopardize the algorithm's convergence. A scaling factor consistent with the algorithm's convergence can be determined according to the following considerations.

The major source for overflows is the sum calculation of the $\beta$ values in line $7$ of Algorithm \ref{alg:linBreg}. Empirical tests conducted based on the testbench developed in [1] indicated that a scaling based on dividing the data vector $\bf y$ by its maximum value allowed to obtain the results shown in Sect. \ref{sect:4} without harming overflow effects. This is due to the firmly non-expansive property \cite{Bauschke2013FixedPointAlg} of the shrink function as well as the negative feedback of the error between $\sum_{s=1}^{k+1} \beta_s^{(i)}$ and $y_k$ (line $7$ of Algorithm \ref{alg:linBreg}). To clarify the negative feedback effect, one could multiply both sides of line 7 by -1, i.e., $-e = y_k - \sum_{s=1}^{k+1} \beta_s^{(i)}$. This procedure would require $\mu_k$, in line 10, to also be multiplied by -1.

The algorithm can be then be interpreted as a stabilizing loop on the values of ${v}_j^{\left(i\right)}$ with the mentioned negative feedback on the deviation between the sum of $\beta$ values (functions of ${v}_j^{\left(i\right)}$) and the corresponding $y_k$, which causes overshoots of the sum of $\beta$ values to be immediately corrected in the next iterations. This leads to the fact that, even in worst case scenarios (multiple up and down trend breaks in the measurements), the sum of $\beta$ values scarcely goes above unit, considering the above mentioned normalization procedure. Moreover, even though its ratio of occurrence is negligible, in the case an overshoot occurs, the excess value would be small thereby not compromising the convergence of the algorithm, as the performance of the quantized version in Sect. {\ref{sect:4}} demonstrates. The closeness of these results to the ones obtained using double precision floating point shows that, practically, harming effects due to overflow can be neglected.

Another important observation is that the stopping criterion developed in \cite{LunglmayrTIM2018} has been presently neglected, and a fixed number of iterations has been considered; the reason for that is twofold. First, the context in \cite{LunglmayrTIM2018} was that of fiber fault location, and the stopping criterion was developed based on a specific phenomenological observation that cannot be extended to more general trend break detection problems. Secondly, with a fixed number of iterations, the standard 64-bit double results can be compared to those of our 20-bit fixed point implementation in a fair scenario. In other words, this quantization could potentially influence the effect of the stopping criterion making the comparison biased.

\section{FPGA Architecture}
\label{sect:3}
\begin{figure*}[ht]
	\centering
	\includegraphics[width=1\textwidth]{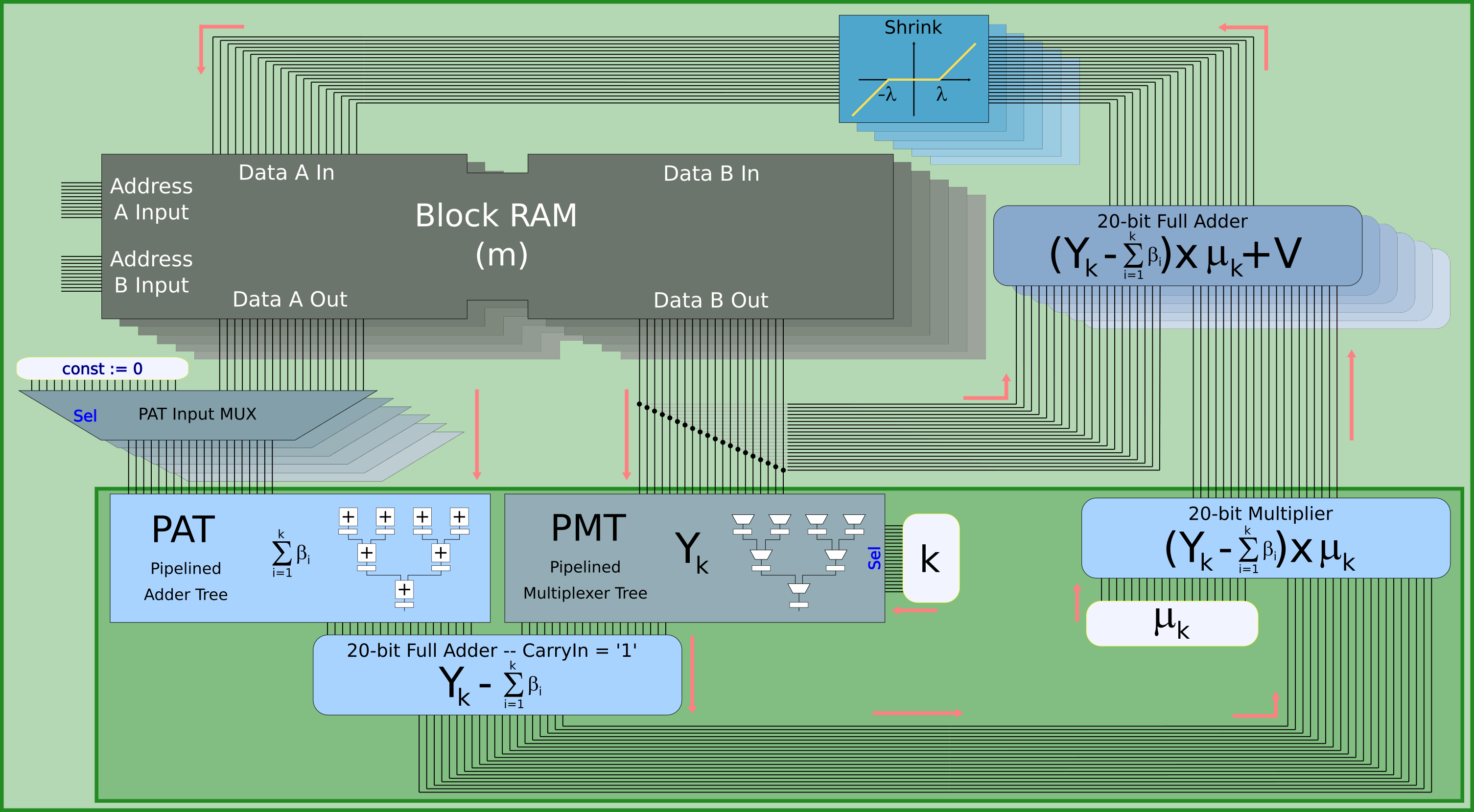}
	\caption{Hardware implementation of a single BRAM slice in the LBI core structure. The parallel structures are pictorially depicted in three-dimensional depth. The Pipelined Multiplexer Tree (PMT) is synchronized to the PAT such that, after a summation, the correct value of $\bf{y}$ is ready for subtraction. The value of $\mu_k$ -- refer to Algorithm \ref{alg:linBreg} -- is calculated in a pipelined CORDIC structure.}
	\label{fig:SingleBRAM}
\end{figure*}

Even though the iterative nature of the Linearized Bregman Iterations algorithm does not allow for parallelization over the iterations, two core operations that permit parallel pipelining can be identified within a single iteration, as presented in Algorithm \ref{alg:linBreg}: the summation of $k$ entries of the vector $\boldsymbol{\beta}$; and the processing (including update, shrinkage, and storage) of vectors $\textbf{v}$ and $\boldsymbol{\beta}$. By instantiating parallel memory structures, both operations, that represent computational bottlenecks of the algorithm's iterations, can be optimized. On one hand, the summation can be efficiently performed in a so-called \textit{parallel adder tree} (PAT) (logarithmic number of time steps) given that the data can be accessed in parallel. On the other, parallel processing of the data in vectors $\textbf{v}$ and $\boldsymbol{\beta}$ can also be accelerated if storage can be performed in parallel. Since the algorithm relies on the computation of several iterations to converge, optimizing these two procedures allows for substantial gains in processing time.

\subsection{Memory Structure}

In order to harness the parallel speedup of the PAT, the entries of vector $\beta$ must also be accessed in parallel, which can be accomplished through the instantiation of parallel Block RAMs (BRAMs). The data storage is structured as follows: 
\begin{align}
\begin{aligned}
&\hspace{0.4cm}\textrm{BRAM}(1) \hspace{0.5cm}\textrm{BRAM}(2) \hspace{1.25cm}\textrm{BRAM}(M)\\
t\bigg\uparrow&\begin{bmatrix}
 \beta[TM+1]        & \beta[TM+2]        & \cdots & \beta[TM+M]        \\
 \vdots             & \vdots             & \cdots & \vdots             \\
 \beta[2M+1]        & \beta[2M+2]        & \cdots & \beta[2M+M]        \\
 \beta[M+1]         & \beta[M+2]         & \cdots & \beta[M+M]         \\
 \beta[1]           & \beta[2]           & \cdots & \beta[M]
\end{bmatrix},\\
&\hspace{3cm}\xrightarrow[m]{\hspace{1cm}}
\end{aligned}
\label{eq:betaMatrix}
\end{align}
where $M$ is the number of parallel BRAMs available in the FPGA. In such a structure, a single arbitrary BRAM, say $m$, will contain entries:
\begin{equation*}
[m+tM] \forall t \in [0;T],m\in[1;M] : T=\ceilingtFrac{\tfrac{N}{M}},
\end{equation*}
where the ceiling operator is denoted as $\ceiling{\cdot}$.

Vector $\beta$, however, is not the only vector stored throughout processing: vectors $\bf{y}$ and $\bf{v}$ are also necessary. Since all these contain the same number $N$ of entries, the data is sectioned such that the address depth of each BRAM is divided in three slices with address pointers (ap) associated with $\beta$ ($\beta_{\textrm{ap}}$), $\bf{y}$ ($\bf{y}_{\textrm{ap}}$), and $\bf{v}$ ($\bf{v}_{\textrm{ap}}$); $\beta_{\textrm{ap}}$ is arbitrarily set to zero. Under this rationale, entries of vectors $\bf{v}$ and $\bf{y}$ would appear at addresses $t+\bf{v}_{\textrm{ap}}$ and $t+\bf{y}_{\textrm{ap}}$, respectively, even though, for simplicity, only entries of vector $\beta$ are shown in Eq. \ref{eq:betaMatrix}. Using this data storage structure, all positions $[tM+1:tM+M]$ of either vectors can be accessed from parallel BRAMs within a clock cycle; such data segment will henceforth be referred to as a \textit{parallel row}, with $t$ the \textit{parallel row} pointer following its definition in Eq.~\ref{eq:betaMatrix}. A block diagram of the digital hardware architecture depicting a single BRAM and including the major structures of the LBI algorithm hardware implementation is presented in detail in Fig.~\ref{fig:SingleBRAM}.

Apart from the PAT, a Pipelined Multiplexer Tree (PMT) is used to select the specific value of the data series $\bf{y}$, namely $y\left(k\right)$, from which the result of the beta sum is subtracted from (refer to line 7 of Algorithm \ref{alg:linBreg}). The architecture of the PMT is such that the number of stages meets that of the PAT, so synchronization between the two outputs is naturally ensured. Furthermore, the selection key that acts on each stage of the PMT is derived from the cyclic iteration index, $k$. The value of $\mu_k = \tfrac{1}{k}$, which involves a computation-heavy division, has been delegated to a pipelined CORDIC structure; the pipeline's stages are pre-filled before the iterations are started and stage propagation is enabled at each new iteration, ensuring that the correct value is always available.

Based on this memory structure, the amount of clock cycles necessary to complete the calculation of $d$ depends both on the number of data points and on the depth of the PAT, which, in turn, depends on the number of instantiated (or available) parallel BRAMs in the hardware structure. For an arbitrary iteration cycle, with cyclic index $k$, the equation that relates these values to the total number of clock cycles is $C'_{\textrm{r}} = \ceilingtFrac{\tfrac{k}{M}} + \ceiling{\log_2M}$, where the subscript refers only to the reading and processing of $\beta$ values up to the output of the PAT. Taking into account also the subsequent subtraction and multiplication steps -- refer to Fig. \ref{fig:SingleBRAM} --, each taking one clock cycle, the total number of clock cycles amounts to $C_{\textrm{r}} = \hat{t} + \lceil\log_2M\rceil + 2$, where $\hat{t} =\ceilingtFrac{\tfrac{k}{M}}$ denotes the maximum value of $t$ during an iteration.

\subsection{PAT Input Control}

Even though a \textit{parallel row} is accessible at each clock cycle due to the parallel instantiation of the BRAMs, clearly not all values in the row will be used during a given iteration with index $k$. For that reason, a multiplexer (\textit{PAT input MUX} in Fig. \ref{fig:SingleBRAM}) is connected immediately after the BRAM ouput with its remaining input connected to a null value. Due to the additive identity property of 0, the output of the multiplexer can be directed to the PAT without corruption of the result while accomodating the parallel storage structure.

\begin{figure}[ht]
	\centering
	\includegraphics[width=1\linewidth]{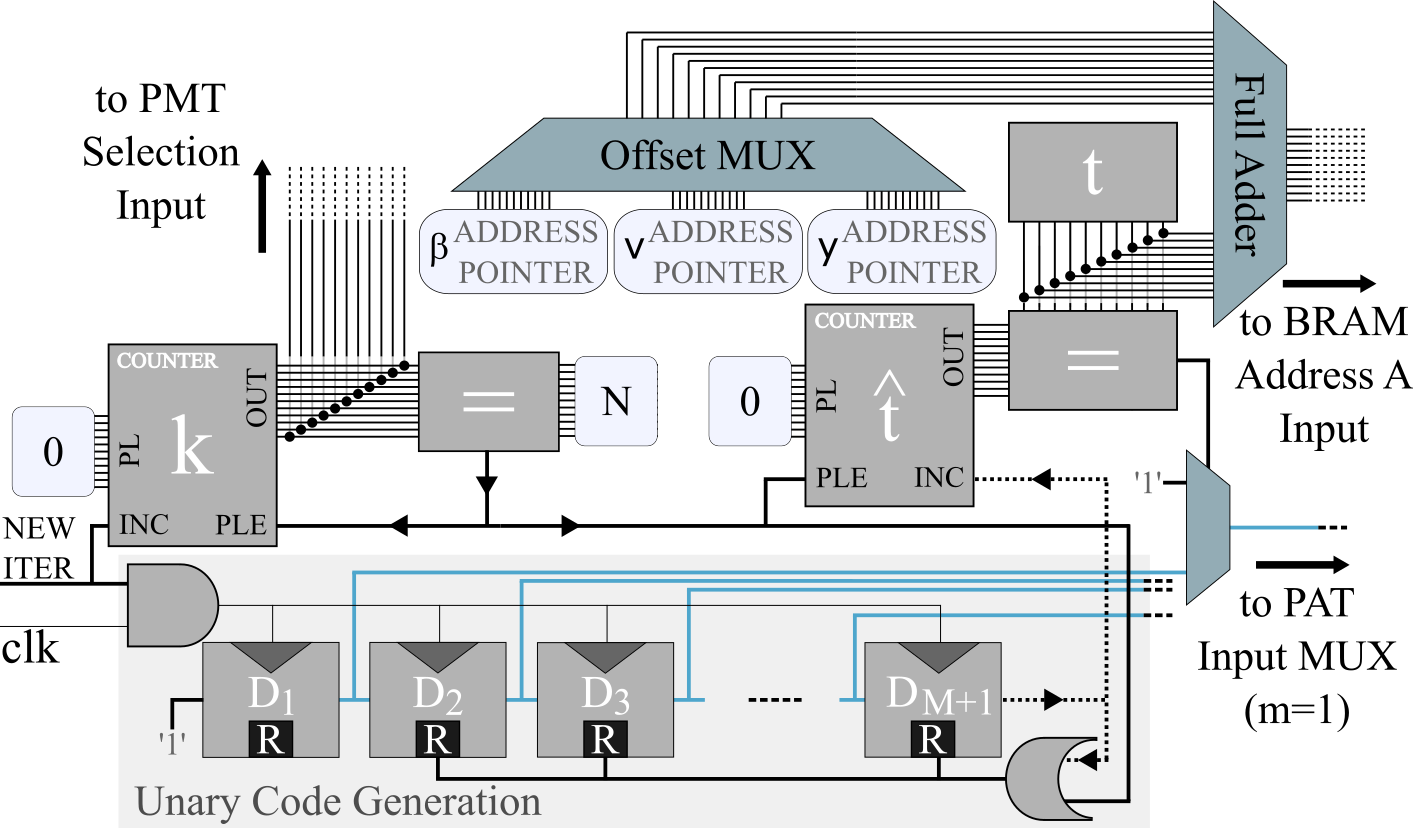}
	\caption{Iteration control architecture. A \textsc{new iter} flag generated by a higher-level unit and the clock signal are the necessary inputs. The cyclic iteration index counter $k$ is implemented through a simple counter with parallel load dependent on the comparison with the signal length $N$. The unary code propagates at each new iteration and, when the $M+1$th stage is reached, it auto-resets while also incrementing the $\hat{t}$ counter. The unary code acts on \textit{PAT input MUX} when $t=\hat{t}$; only the selection for $m=1$ is depicted for clarity. The different address pointers are combined with the counter $t$ to produce the correct BRAM address.}
	\label{fig:PAT_ctrl_Arch}
\end{figure}

The selection signal that controls the \textit{PAT input MUX} is derived based on the fact that replacing BRAM outputs by 0 is only necessary during the last \textit{parallel row} access, i.e., when $t=\ceilingtFrac{\tfrac{k}{M}}=\hat{t}$. Selection is thus based on an auxiliary counter that records the aforementioned value and on a so-called unary code (or thermometer code), which encodes the last column index that contributes to the sum. Fig. \ref{fig:PAT_ctrl_Arch} depicts the control unit responsible for handling the BRAM input address and selection of PAT input.

\subsection{$\boldsymbol{\beta}$ and \textbf{\textrm{v}} Storage}

An indispensible step of the algorithm is the correct storage of the vectors $\boldsymbol{\beta}$ and $\bf{v}$ after processing. According to Algorithm \ref{alg:linBreg}, all the elements of vector $\boldsymbol{\beta}$ are processed by the shrink function right after processing of the vector $\bf{v}$. As previously pointed out, acceleration of the storage procedure tackles one of the algorithm's bottlenecks: the number of clock cycles taken by the storage procedure scales the total number of clock cycles necessary for the algorithm to elapse. Both the fact that the BRAMs allow for writing and reading from two independently addressed ports and that if $\lambda$ is set to zero in the shrink function it implements the identity transformation have been harnessed to perform data storage optimization, as it is detailed as follows.

One of the BRAM's ports (taken as B without loss of generality in Fig. \ref{fig:SingleBRAM}) is responsible for reading the values of $v$ from the memory while the other port (A) is responsible for storing the values of $\beta$ and $v$. The addresses are controlled such that, on the first clock cycle, values of $v$ in a \textit{parallel row} are read (through port B), processed in the 20-bit full adder, and sent to the shrink function with $\lambda = 0$. Therefore, at the following clock cycle, the stable value of $v$ can be stored (through port A) at the same time as the value of $\lambda$ is changed in the shrink function and processes the values of $v$ being read (through port B). In the third clock cycle, a stable value of $\beta$ is stored (through port A) while the values of $v$ from the following \textit{parallel row} are accessed (through port B), initiating a new storage cycle for a subsequent \textit{parallel row}.

\subsection{Total Clock Cycle Estimation}

The net amount of clock cycles per \textit{parallel row} storage is, thus, two if one does not compute the very first and last accesses; therefore, the number of clock cycles necessary at an arbitrary iteration with cyclic index $k$ is $C_{\textrm{s}} = 2\ceilingtFrac{\frac{k}{M}}+2=2\hat{t}+2$. Two extra clock cycles are also necessary for the hand-shaking protocol between the iteration control unit (presented in Fig. \ref{fig:PAT_ctrl_Arch}) -- whose control over the BRAMs address is releaved -- and the writing unit that takes over control and stores vectors $\beta$ and $v$, i.e., $C'_{\textrm{s}} = 2\hat{t}+4$. Combining this value with the previously determined $C_{\textrm{r}}$ for the reading and processing of $\beta$ values in the PAT, the total number of clock cycles spent in an arbitrary iteration with index $k$ is $C_{\textrm{T}}=3(\hat{t} + 2) + \ceiling{\log_2M}$. 

The total number of clock cycles taken by the algorithm to elapse can be easily derived from this equation by summation over L, the total number of iterations:
\begin{equation}
C\!=\!F\!+\!\sum_{i=1}^{L}\left[3\left(\ceilingFrac{\frac{\left(\left(i-1\right)\% N\right)+1}{M}}\!+\!2\!\right)\!+\!\ceiling{\log_2 M}\right].
\label{eq:totalCC}
\end{equation}
The factor $F$ in Eq. \ref{eq:totalCC} accounts for pre- and post-processing instructions performed by the control unit such as: master resets; granting control over the BRAMs; and, most importantly, preemptively filling up the pipelined CORDIC that calculates $\mu_k$. However, as will be described in the next Section, the value of $F$ is much smaller than the total number of clock cycles taken by the core procedure.

\begin{figure}[ht]
	\centering
	\includegraphics[width=1\linewidth]{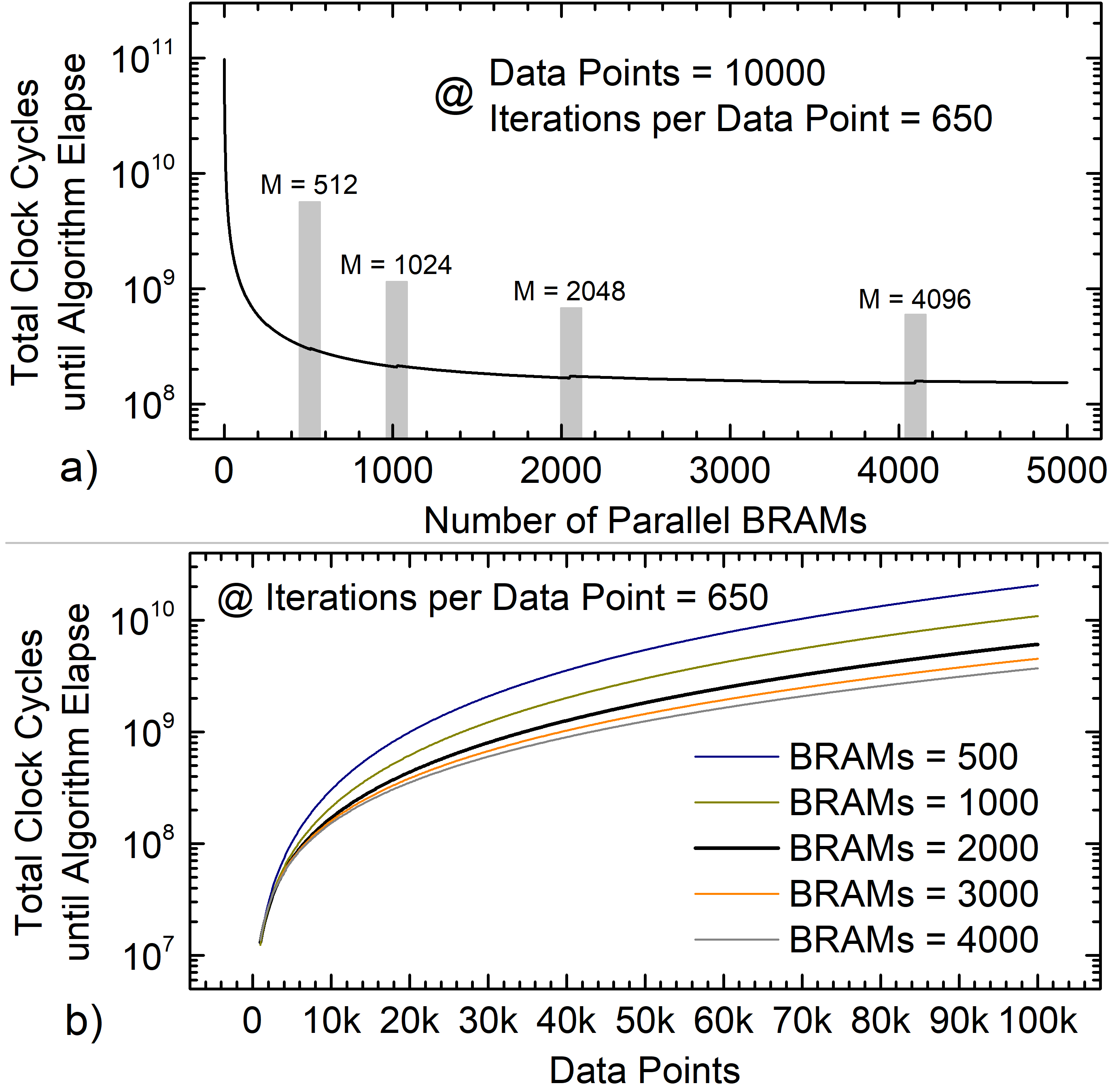}
	\caption{Estimate of total number of clock cycles necessary for the algorithm to elapse considering the presented architecture. (a) Dependence with respect to number of available parallel BRAMs for a fixed number of 10000 data points and 650 iterations per data point. The gray-shaded areas highlight the transition between powers of 2, which manifests as sharp increases in the calculated value of $C$. (b) Dependence with respect to the number of data points for a fixed number of available BRAMs and 650 iterations per data point. The highlighted curve corresponds to 2000 BRAMs, which is the maximum available for the largest target FPGA studied here.}
	\label{fig:clock_cycle}
\end{figure}

Fig. \ref{fig:clock_cycle} presents the dependence of the total number of clock cycles until the algorithm elapses with both the number of available parallel BRAMs for a fixed number of data points and with the number of data points for a fixed number of available BRAMs. In both cases, the iterations per data point (defined as $L/N$) is fixed at 650, a realistic value which will be discussed further in the next Section. Considering a realistic scenario, where the maximum clock frequency achievable in the target FPGA is around 100 MHz, a 10000-point data series would be processed in less than two seconds, which represents an approximately 100 gain factor when compared to the Julia implementation reported in \cite{LunglmayrTIM2018}.

\section{Validation and Synthesis}
\label{sect:4}
Comparison between the software-defined hardware implementation of the Linearized Bregman Iterations algorithm using the architecture presented in the previous Section and its software implementation counterpart \cite{LunglmayrTIM2018} permits validating the former. In order to provide a bit-true validation, the \textit{SFIXED} standard used in the VHDL simmulation was implemented in Julia allowing one to accompany, step-by-step, the evolution of the algorithm on both platforms and identify any discrepancies. Due to the fact that the rounding procedure is the same for both, no such discrepancies were observed; the fixed point Julia simulation code outputs \emph{exactly} the same values as of the hardware implementation. The validation of the hardware implementation and the demonstration of its equivalence to the Julia \textit{SFIXED} implementation creates a versatile tool to estimate the performance of the FPGA results on a software environment.

For the simulation of the hardware implementation, the \textsc{ModelSim} VHDL simulation environment was employed. In such an environment, both the evolution of the algorithm as well as the number of clock cycles necessary to run each iteration can be extracted, so the results of Eq. \ref{eq:totalCC} can also be ascertained. Even though an extremely reliable and versatile tool, VHDL simulation offers a drawback in terms of running time: simulating a high number of BRAMs or a large dataset can be extremely time-consuming. For this reason, a predetermined set of parameters (data points, iterations, and number of BRAMs) were chosen to showcase the validity of the hardware implementation.

Table \ref{tab:validation} contains the information regarding the simulation of the hardware structure under the different parameter conditions, where B stands for the number of BRAMs, and L and N follow the previously defined notation. The estimated number of clock cycles based on Eq. \ref{eq:totalCC} that appear in Table \ref{tab:validation} take into account the required $F=21$ extra clock cycles for initialization and control. The asterisk in the last column indicates that 2048 BRAMs is actually above the 2000 maximum available number of BRAMs with 20 bit-wide data entries in the target \textsc{Altera Stratix V} FPGA.

\begin{table}[ht]
\centering
\caption{VHDL Simulation -- Validation}
\setlength\tabcolsep{3pt} 
\setlength\extrarowheight{6pt}
\begin{tabular}{c|ccc|ccc|ccc|ccc|ccc}
 & \rotatebox{90}{N = 10} & \rotatebox{90}{B = 4} & \rotatebox{90}{L = 10} &\rotatebox{90}{N = 10} & \rotatebox{90}{B = 4} & \rotatebox{90}{L = 100} & \rotatebox{90}{N = 100} & \rotatebox{90}{B = 4} & \rotatebox{90}{L = 100} & \rotatebox{90}{N = 100} & \rotatebox{90}{B = 4} & \rotatebox{90}{L = 1000} & \rotatebox{90}{N = 1000} & \rotatebox{90}{B = 4} & \rotatebox{90}{L = 1000} \\
\hline\hline
$C$ (Eq. \ref{eq:totalCC}) & \multicolumn{3}{c|}{155} & \multicolumn{3}{c|}{1361} & \multicolumn{3}{c|}{4721} & \multicolumn{3}{c|}{47021} & \multicolumn{3}{c}{384521} \\
Clk. Cyc. & \multicolumn{3}{c|}{155} & \multicolumn{3}{c|}{1361} & \multicolumn{3}{c|}{4721} & \multicolumn{3}{c|}{47021} & \multicolumn{3}{c}{384521} \\
\hline
 & \rotatebox{90}{N = 1000} & \rotatebox{90}{B = 128} & \rotatebox{90}{L = 1000} &\rotatebox{90}{N = 5000} & \rotatebox{90}{B = 1024} & \rotatebox{90}{L = 5000} & \rotatebox{90}{N = 10000} & \rotatebox{90}{B = 1024} & \rotatebox{90}{L = 10000} & \rotatebox{90}{N = 15000} & \rotatebox{90}{B = 1024} & \rotatebox{90}{L = 15000} & \rotatebox{90}{N = 15000 } & \rotatebox{90}{B = 2048$^{\ast}$} & \rotatebox{90}{L = 15000} \\
\hline\hline
$C$ (Eq. \ref{eq:totalCC}) & \multicolumn{3}{c|}{26269} & \multicolumn{3}{c|}{124301} & \multicolumn{3}{c|}{321781} & \multicolumn{3}{c|}{592461} & \multicolumn{3}{c}{442989} \\
Clk. Cyc. & \multicolumn{3}{c|}{26269} & \multicolumn{3}{c|}{124301} & \multicolumn{3}{c|}{321781} & \multicolumn{3}{c|}{592461} & \multicolumn{3}{c}{442989} \\
\end{tabular}
\label{tab:validation}
\end{table}

The results of Table \ref{tab:validation} are in excellent agreement with the expectations, which translate into: validation of the hardware implementation as well as demonstration of its equivalence to the Julia \textit{SFIXED} software implementation; and verification of the validity of Eq. \ref{eq:totalCC}, which, in turn, is a validation of the results of Fig. \ref{fig:clock_cycle}. The concluding step of this Section is, then, to synthesize the hardware so that the maximum achievable clock frequency can be extracted. As previously commented, the clock frequency, combined with the total number of clock cycles necessary for the algorithm to elapse, can be used to estimate the amount of time the algorithm will take to execute. Furthermore, as a by-product of the synthesis results, it is possible to assess the percentage of FPGA resources occupied by the architecture, which, in turn, provides the means for selecting the target FPGA for hardware implementation. The results are summarized in Table \ref{tab:synth_n_timing}, where the \textsc{Intel Quartus Prime} synthesis software was used.

\begin{table*}[ht]
\centering
\caption[caption]{Target FPGAs Synthesis and Timing Results\\\vspace{-0.15cm}\hspace{\textwidth}\textrm{Processing times calculated for $N = 10000$ and with $L = 6.5 \times 10^6$}}
\setlength\tabcolsep{3pt} 
\setlength\extrarowheight{6pt}
\begin{tabular}{c|ccccc|ccccc}
 & \multicolumn{5} {c} {Stratix V: 5SGSMD5K2F40C2}  & \multicolumn{5} {c} {Cyclon V: 5CSXFC6D6F31C6}\\
\hline
BRAMs & ALMs	& Registers	&Memory [Bits] & Max. Clk. & Proc. & ALMs	& Registers	&Memory [Bits] &  Max. Clk. & Proc. \\
 & out of $172,\!600$ & & out of $41,\!246,\!720$ & Freq. [MHz] & Time [s] & out of $41,\!910$ & & out of $5,\!662,\!720$ & Freq. [MHz] & Time [s] \\
\hline
$1024 $ & $ 135,\!571 ( 79 \% )  $ & $ 88,\!938 $ & $ 20,\!971,\!520 ( 51 \% ) $ & 109.9 & 1.91 & $-^{\textrm{1}}$ & $-^{\textrm{1}}$ & $-^{\textrm{1}}$ & $-^{\textrm{1}}$ & $-^{\textrm{1}}$ \\
$512 $   & $ 68,\!135 ( 39 \% ) $ & $ 45,\!595 $ & $ 10,\!485,\!760 ( 25 \% )$ & 166.97 & 1.78  & $-^{\textrm{1}}$ & $-^{\textrm{1}}$ & $-^{\textrm{1}}$ & $-^{\textrm{1}}$ & $-^{\textrm{1}}$ \\
$256 $   & $ 36,\!098( 21 \% ) $ & $ 25,\!147 $ & $ 5,\!242,\!880 ( 13 \% )$ & 173.28 & 2.78 & $36,\!548 ( 87 \% )$ & $24,\!741$ & $2,\!621,\!440 ( 46 \% ) $ & 81.13 & 5.94\\
$128 $   & $ 19,\!196 ( 11 \% ) $ & $ 13,\!839 $ & $ 2,\!621,\!440 ( 6 \% )$ & 182.12 & 4.70 & $20,\!178 ( 48 \% )$ & $13,\!475$ & $1,\!310,\!720 ( 23 \% )$ & 95.37 & 8.98\\
$16 $     & $ 4,\!982 ( 3 \% ) $ & $ 4,\!012 $ & $ 327,\!680 ( < 1 \% )$ & 182.35 & 33.83 & $5,\!000 ( 12 \% )$ & $3,\!928$ & $163,\!840 ( 3 \% )$ & 92.94 & 66.37\\
$4 $       & $ 3,\!435 ( 2 \% )  $ & $ 2,\!925 $ & $ 81,\!920 ( < 1 \% ) $ & 187.86 & 130.08  & $3,\!427 ( 8 \% )$ & $2,\!929$ & $40,\!960 ( < 1 \% )$ & 91.73 & 266.40
\end{tabular}
\label{tab:synth_n_timing}
\begin{flushleft}
$^{\textrm{1}}$ : Design too large to fit into device.
\end{flushleft}
\end{table*}

Up to 1024 BRAMs could be instantiated in the \textsc{Stratix V}, with as high as 109 MHz maximum clock frequency yielding a 1.91 second processing time for 10000-long data series considering 650 iterations per sample (overall, 6.5 million iterations). This result represents a 100 speedup factor in processing time with respect to software implementations under the same data conditions but running on a \textsc{Intel Xeon CPU E5-2690 v4} at 2.6 GHz and 512 GB RAM, a major achievement, which advocates for the dedicated hardware solution for trend break detection problem. It is also interesting to note that, for a smaller FPGA, the \textsc{Cyclon V}, a $\sim 16$ gain factor with respect to the software implementation was achieved, which is interesting in the sense that smaller FPGAs exhibit, generally, significantly lower costs, but could still deliver processing times in the range of a few seconds.

The results from Table \ref{tab:synth_n_timing} also indicate that a compromise between instatiation of higher number of BRAMs (which reduces the total number of clock cycles necessary for the algorithm to elapse as determined by Eq. \ref{eq:totalCC}) and the maximum achievable clock frequency exists. In fact, the processing time for 512 instantiated BRAMs was lower than that of 1024 because the gain in clock frequency superseeded that of the reduction of clock cycles. It should be mentioned, however, that the place and routing problem is an extremely complex one and the algorithms that solve it may not always reach the best possible solution, so the clock frequency values obtained should be interpreted as lower bounds. Finally, to put the results into an application proned perspective, fiber profiles as long as 50 km could be analyzed in search for breaks in under 10 seconds \cite{LunglmayrTIM2018}.

\section{Case Study Results}
\label{sect:5}
\begin{figure}[ht]
	\centering
	\includegraphics[width=1\linewidth]{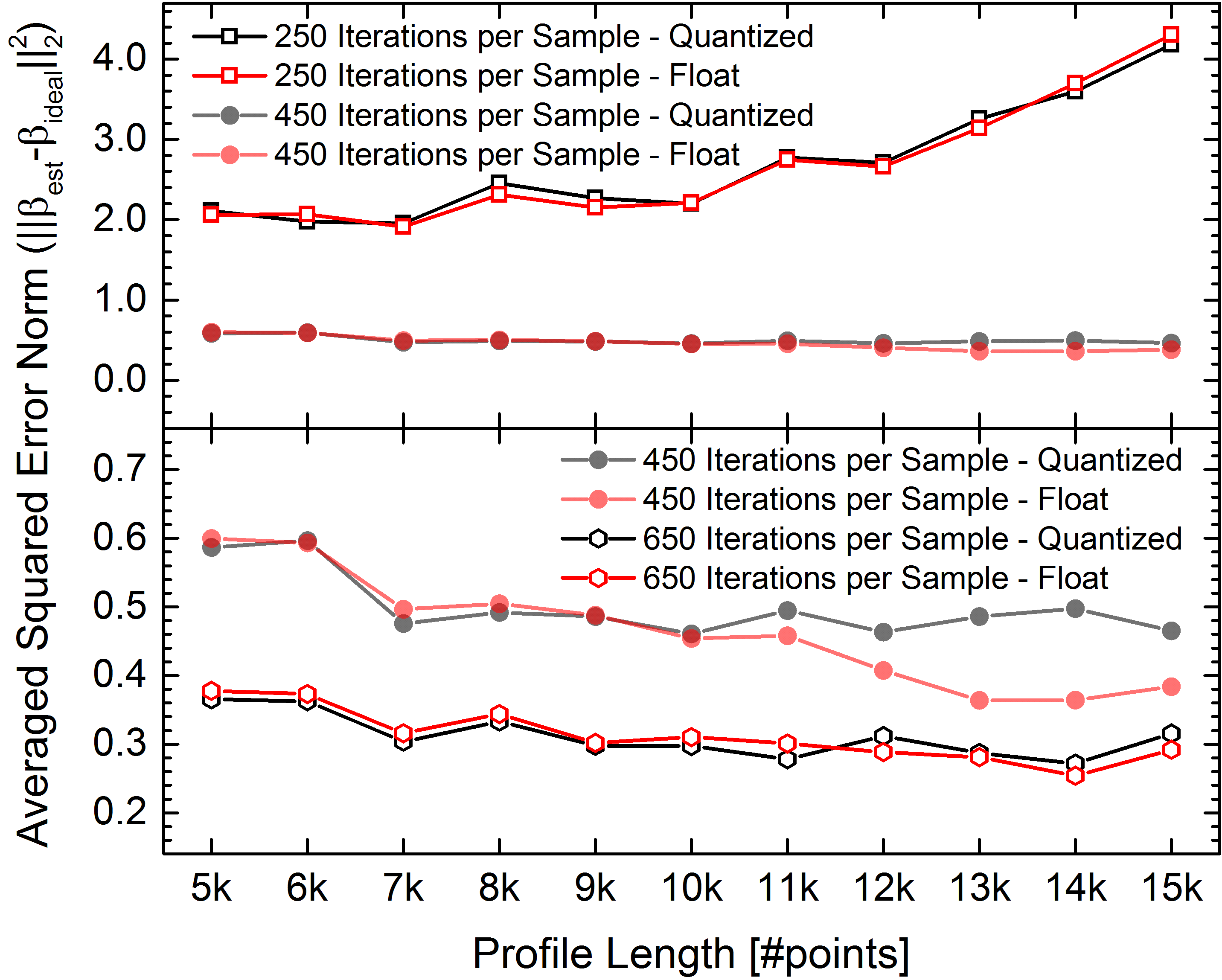}
	\caption{Estimation results for both \textsc{SFIXED} and floating point implementations for different profile lengths and different numbers of iterations per sample; L is simply this value multiplied by the number of points (N) in the profile. The results are divided into two panels for ease of visualization.}
	\label{fig:quant_impl}
\end{figure}

Validation of the software Julia \textit{SFIXED} implementation performed in Section \ref{sect:4} allows one to investigate aspects of the hardware implementation in a more suitable simulation environment. This is important due to the amount of simulation workload necessary to yield statistically relevant results. Point in fact is the investigation of the quality of estimation as L, the total number of iterations, increases, for which the results of over 15000 different testbench simulated profiles, analyzed with the hardware-validated bit-true Julia \textit{SFIXED} implementation, is presented in Fig. \ref{fig:quant_impl}.

The objective of this extensive study is twofold: firstly, as it was already mentioned, to empirically determine the interval of number of iterations per sample for which the estimation quality reaches a reliable level; and, secondly, to compare the quality of estimation between the 64-bit double implementation, and the 20-bit \textit{SFIXED} implementation for profiles with different number of data points and with different number of iterations per sample. The profiles analyzed with the two versions of the algorithm (64-bit double and 20-bit \textit{SFIXED}) were created following the rationale of the testbench profile generation presented in \cite{LunglmayrTIM2018}, i.e., noiseless sparse vectors with randomly sorted magnitudes ($\beta_{\textrm{ideal}}$) are used to create profiles using the candidate matrix \textbf{A}, to which white gaussian noise is added. The result of the estimation is $\beta_{\textrm{est}}$, which is then compared to $\beta_{\textrm{ideal}}$ using the squared error norm; the closer to zero error between the estimated and ideal vectors, the better is the estimation. This procedure is similar, but not equivalent, to the one described in \cite{LunglmayrTIM2018} since no slopes are considered and the noise parcel is not derived based on any phenomenological observations.

From the results of Fig. \ref{fig:quant_impl}, it is possible to conclude that the differences in estimation between the \textit{SFIXED} and 64-bit double implementations are negligible, i.e., the hardware implementation will have no problems achieving comparable estimation accuracy as the software version, e.g. in \cite{LunglmayrTIM2018}. Furthermore, it becomes clear that, after 450 iterations per sample, the accuracy stabilizes, with a averaged squared error norm value in the order of 0.5. This indicates that the results from Fig. \ref{fig:clock_cycle} with 650 iterations per sample are indeed realistic. Finally, this result is also useful when interpreted along with those of Fig. \ref{fig:clock_cycle}: a compromise between the total number of clock cycles before the algorithm elapses and the quality of the estimation can be found and, in specific cases, one of these can be sacrificed (increasing the processing time or allowing for a worst estimate) to boost the other (faster results or extremely precise estimation).

\section{Conclusions}
\label{sect:6}

Trend break detection, or level-shift detection, is a problem that permeates several science fields, and an efficient, accurate, and highly reliable processing unit to solve it is desirable. Combining the flexible hardware design tools of Field Programmable Gate Arrays and the efficient Linearized Bregman Iterations algorithm allowed the development of such unit. The manipulation of the data storage structure as well as the algorithm flow and control in hardware yielded an up to 100 times gain in processing time when compared to a personal computer while maintaining all the observed qualities of the algorithm, such as low estimation error and high level-shift detection precision.

The proposed hardware architecture can be implemented in different sized-FPGAs, with the main distinctions being the amount of available dual-block RAMs and maximum achievable clock frequency, characteristics which are hardware-dependent. On a middle-sized chip such as the \textsc{Altera Cyclone V}, the hardware supports up to 256 parallel BRAMs with a maximum clock frequency of 81 MHz and a total processing time of 6 seconds. Such processing prowess can be directed towards on-line data supervision such as optical fiber monitoring, which constitutes an exciting future point of investigation. Furthermore, incorporating advanced signal processing techniques into the the hardware design in order to eliminate any pre-processing step while increasing the convergence speed is also a sought-after goal for future studies.

\section*{Acknowledgment}
Financial support from brazilian agency CNPq is acknowledged by F. Calliari. This work has been supported by the COMET-K2 ``Center for Symbiotic Mechatronics'' of the Linz Center of Mechatronics (LCM) funded by the Austrian federal government and the federal state of Upper Austria.

\bibliographystyle{IEEEtran}
\bibliography{References}

\end{document}